\documentclass[11pt,showkeys,showpacs,superscriptaddress,nofootinbib,floatfix]{revtex4-2}

\usepackage[T1]{fontenc}
\usepackage[english]{babel}
\usepackage{microtype}
\usepackage{physics}

\usepackage{newtxtext}
\usepackage{newtxmath}

\usepackage{bm}
\usepackage{slashed}

\usepackage[dvipsnames,table,xcdraw]{xcolor}
\definecolor{lightyellow}{RGB}{255,250,205}

\usepackage{graphicx}
\usepackage{epstopdf}

\usepackage{array}        
\usepackage{tabularx}
\usepackage{multirow}
\usepackage{longtable}

\usepackage{url}
\usepackage{orcidlink}

\usepackage{hyperref} 
\hypersetup{ 
  colorlinks=true,
  linkcolor=blue,  
  citecolor=cyan,
  urlcolor=blue,           
}

\begin{document}

\title{Revisiting relativistic corrections to inclusive $J/\psi$ production at B factories: Complete expansion for three-body quarkonium production}

	\author{Sheng-Juan Jiang $^{(a)}$}
	\author{Sai Cui $^{(a)}$}
	\author{Guang-Zhi Xu $^{(a)}$}
	\email{xuguangzhi@lnu.edu.cn}
	\author{Kui-Yong Liu $^{(b,a)}$}
	\email{liukuiyong@lnu.edu.cn}
	\affiliation{ {\footnotesize (a)~School of Physics, Liaoning University, Shenyang 110036, China}\\
		{\footnotesize (b)~School of Physics and Electronic Technology, Liaoning Normal University, Dalian 116029, China}}

\begin{abstract}
We revisit the calculations of relativistic corrections to inclusive $J/\psi$ production at B factories. For quark-level subprocesses with three-body final states, we carry out a full expansion of the final-state kinematic parameters. The resulting cross sections are theoretically self-consistent and independent of the choice of integration variables. In the high-energy limit, both cross-section magnitudes and the line shapes of correction curves for energy and momentum distributions agree remarkably well with fragmentation-function calculations. We find that $\mathcal{O}(v^2)$ corrections suppress the cross section by roughly $14.55\%$ in the $J/\psi + X_{c\bar{c}}$ channel and enhance it by approximately $12.18\%$ for the $J/\psi + X_{\text{non-}c\bar{c}}$ channel. 
After incorporating published $\mathcal{O}(\alpha_s)$ corrections, color-octet contributions, two-photon production channels, and feed-down effects, tension persists between theoretical predictions and experimental measurements. 
This indicates that complete prompt-quarkonium calculations through $\mathcal{O}(\alpha_s, v^2)$, or evaluations incorporating higher-order corrections, are required to resolve this tension.
\end{abstract}


\date{\today}

\maketitle

\section{Introduction}
The production of heavy quarkonia, in particular $J/\psi$, in $e^+e^-$ annihilation provides a stringent testing ground for nonrelativistic QCD (NRQCD) factorization~\cite{Bodwin:1994jh,QuarkoniumWorkingGroup:2004kpm}.
Over the past two decades, inclusive and exclusive $J/\psi$ production at $B$ factories with $\sqrt{s}\simeq10.58~\mathrm{GeV}$ has provided a clean platform for probing both perturbative and nonperturbative mechanisms underlying quarkonium formation \cite{Liu:2002wq,Braaten:2002fi,Zhang:2005cha,Gong:2007db,Huang:2022dfw,Feng:2019zmt,He:2007te,Liu:2003jj,Liu:2003zr,Zhang:2006ay,He:2009uf,Ma:2008gq,Gong:2009kp,Hagiwara:2004pf,Chen:2022qli,Shao:2014rwa,He:2009uf}.
The measurements from the Belle and BaBar Collaborations have shown that both the double-charm channel $e^+e^- \to J/\psi + c\bar c + X$ and the non-$c\bar c$ channel $e^+e^- \to J/\psi + X_{\rm non-c\bar c}$ contribute sizably to the total inclusive $J/\psi$ production rate~\cite{Pakhlov:2004au,BaBar:2005nic,BaBar:2001lfi,Belle:2002tfa,Belle:2001lqi}. The latest data for the prompt $J/\psi$ production cross sections are presented separately for $c \bar{c}$ and non-$c \bar{c}$ final states~\cite{Belle:2009bxr}:
\begin{align}
\sigma[e^+ e^- \to J/\psi + X_{c\bar{c}}] &= 0.74 \pm 0.08^{+0.09}_{-0.08}~\text{pb}, \\
\sigma[e^+ e^- \to J/\psi + X_{\text{non}-c\bar{c}}] &= 0.43 \pm 0.09 \pm 0.09~\text{pb}.
\end{align}
However, the leading-order (LO) NRQCD predictions significantly underestimated these measurements, indicating the necessity of higher-order corrections \cite{Liu:2003jj}.
Next-to-leading-order (NLO) QCD corrections significantly increase the cross section of the $J/\psi+X_{c\bar{c}} $ channel \cite{Zhang:2006ay,Gong:2009ng}, whereas their effect on the $J/\psi+X_{\text{non}-c\bar{c}} $ channel is comparatively moderate, yielding an overall enhancement of roughly $20\%$ \cite{Ma:2008gq,Gong:2009kp}.
This strong channel-dependent behavior refines the theoretical prediction for the ratio of the $J/\psi+X_{c\bar{c}}$ cross section to the total inclusive $J/\psi$ cross section and partially eases the theory–experiment tension~\cite{Zhang:2006ay,Ma:2008gq,Gong:2009kp,Gong:2009ng}.

Nevertheless, this agreement is imperfect and is sensitive to the renormalization scale. Furthermore, existing theoretical analyses include color-octet contributions for the $X_{c\bar{c}}$ channel yet neglect those for the $X_{\text{non-}c\bar{c}}$ channel. These neglected color-octet corrections are sizable if one employs color-octet long-distance matrix elements (LDMEs) extracted from NLO fits to hadron-collider data. As a result, residual tension remains between theory and experiment after all relevant contributions are incorporated.

The relativistic corrections may also play an important role in achieving a more complete description of inclusive $J/\psi$ production. 
Previous studies reported that relativistic corrections to the $J/\psi+X_{c\bar{c}}$ channel are extremely small, below $0.5\%$ \cite{He:2007te}, while they enhance the cross section of the $J/\psi+X_{\text{non-}c\bar{c}}$ channel by approximately $20\%$ to $30\%$ \cite{He:2009uf}. Including such corrections further widens the tension between theoretical predictions and experimental data. 
On the other hand, these prior calculations neglect relativistic phase-space effects and $\mathcal{O}(v^2)$ dependence of kinematic parameters associated with final-state momenta in the scattering amplitudes, owing to the complexity of three-body quark production processes.
An alternative computational scheme was proposed in Ref.~\cite {Jia:2010fw} for the $X_{\text{non-}c\bar{c}}$ channel, where the heavy quark mass is expanded around the physical quarkonium mass. This treatment deviates from the conventional calculation strategy.
In the present work, we revisit the magnitude of these relativistic corrections within the conventional computational strategy. We first analyze the $\mathcal{O}(v^2)$ dependence of final-state kinematic parameters, whose contributions are critical for a complete theoretical description. By incorporating these terms, we obtain the complete set of relativistic corrections for three-body $J/\psi$ production.

The rest of this paper is organized as follows. The theoretical framework is presented in Sec.~\ref{sec:calculation}, followed by the numerical results and discussions in Sec.~\ref{sec:results}. Finally, Sec.~\ref{sec:Conclusion} summarizes this work.

\section{Theoretical framework}\label{sec:calculation}

Within the NRQCD factorization framework, the inclusive differential cross section for $e^+e^- \to J/\psi + X$, including the relativistic corrections, factorizes into short-distance coefficients (SDCs) and long-distance matrix elements (LDMEs) as
\begin{equation}
\mathrm{d}\sigma(J/\psi) = \frac{F}{m_c^2}\langle0|\mathcal{O}^{J/\psi}|0\rangle+\frac{G}{m_c^4}\langle0|\mathcal{P}^{J/\psi}|0\rangle,
\label{eq:sig}
\end{equation}
where $\langle0|\mathcal{O}^{J/\psi}|0\rangle$ and $\langle0|\mathcal{P}^{J/\psi}|0\rangle$ denote the LO and $\mathcal{O}(v^2)$ LDMEs, respectively.
The corresponding SDCs $F, G$ are determined by matching perturbative QCD (pQCD) and perturbative NRQCD calculations for the hard production of the constituent $c\bar{c}$ pair.
In this work, we adopt the color-singlet framework to perform the pQCD calculation.
The differential cross section is given by the phase-space integral of the squared scattering amplitude.
\begin{equation}
\begin{aligned}
& \mathrm{d}\sigma\left[c\bar{c}(^3S_1^{1})\right] = \frac{1}{2s} \mathrm{d}\Phi \overline{\sum} \left| \mathcal{M}
\left[
e^+e^- \to c\bar{c}(^3S_1^{1}) + X
\right]
\right|^2 .
\end{aligned}
\end{equation}
Here $1/(2s)$ denotes the Lorentz-invariant flux factor, while $s$ stands for the squared center-of-mass (CM) collision energy, and the sum $\overline{\sum}$ averages over initial-state spins and sums over final-state spins. The scattering amplitude $\mathcal{M}$ is constructed as follows
\begin{equation}\label{eq:amp}
\begin{aligned}
& \mathcal{M}
\left[
e^+e^- \to c\bar{c}(^3S_1^{1}) + X
\right] =
\sum_{s \bar{s},i j}
\langle\tfrac{1}{2}, s; \tfrac{1}{2}, \bar{s} \mid 1, S_{z}\rangle 
\langle 3 i; 3 \bar{j} \mid 1 \rangle
\\
&\quad \times
\mathcal{A}(e^+e^- \to c_{s,i} + \bar{c}_{\bar{s},\bar{j}} +X).
\end{aligned}
\end{equation}
We adopt the Lorentz-covariant spinor projection technique to project the $c\bar{c}$ pair onto the spin-triplet $^3S_1^1$ state~\cite{Berger:1980ni,Bodwin:2002cfe,Keung:1982jb,Braaten:1996jt}. The corresponding Lorentz-covariant spin and color projection operators are written as
\begin{equation}\label{eq10b}
\begin{aligned}
&\sum_{s \bar{s}}
v(q,\bar{s})\,\overline{u}(q,s)
\left\langle \tfrac{1}{2},s;
\tfrac{1}{2},\bar{s} \middle| 1,S_{z} \right\rangle
\\[4pt]
&\quad=
-\frac{1}{2\sqrt{2}(E_q+m_c)}
\left(\slashed{p}_{\bar{c}}-m_c\right)
\slashed{\epsilon}(S_{z})
\frac{\slashed{P}+2E_q}{2E_q}
\left(\slashed{p}_c+m_c\right),\\
&\sum_{i j}
\langle 3 i; 3 \bar{j} \mid 1 \rangle = \frac{\delta_{ij}}{\sqrt{3}}.
\end{aligned}
\end{equation}
In the above form, we use the standard Dirac spinor normalization $\bar{u}u = 2m_c$ and $\bar{v}v = -2m_c$. $\epsilon$ is the spin polarization vector.
We denote the four-momenta of the heavy quark and antiquark as $p_c$ and $p_{\bar{c}}$, respectively. 
Their decomposition in an arbitrary inertial frame takes the form~\cite{He:2007te,Braaten:1996jt,Ma:2012ex}:
\begin{equation}
\begin{aligned}
p_c &= \frac{1}{2}P + q,\quad
p_{\bar{c}} &= \frac{1}{2}P - q.
\end{aligned}
\end{equation}
Here $P$ is the meson four-momentum and $q$ is the relative four-momentum of the heavy quark–antiquark pair. In the meson rest frame, $P=(2E_q,\bm 0)$ and $q=(0,\bm q)$, with $E_q = \sqrt{m_c^2 + |\bm q|^2}$. The magnitude $|\bm q|$ is Lorentz invariant.
Relativistic corrections are systematically evaluated by expanding the amplitude in powers of $|\bm q|$. Both $E_q$ and $q$ enter the amplitude explicitly, and additional kinematic variables also carry $|\bm q|$ dependence. 
These kinematic quantities can be written as functions of $E_q$ and angular variables (the scattering angles of final-state particles measured in the initial or final CM frame). In the standard expansion procedure, these angular degrees of freedom are treated as independent of $|\bm q|$.
Furthermore, we note that these angular variables are Lorentz scalars, which ensures manifest covariance order-by-order in the amplitude expansion.
In fact, such expansion procedures have been widely adopted in prior works on relativistic corrections to quarkonium decays and two-body production processes \cite{Li:2013csa,Xu:2012am,He:2014sga,He:2007te,He:2009uf,Berezhnoy:2003hz,Bodwin:2003wh}.
In particular, a universal coefficient is introduced for the final-state momenta in heavy quarkonium decays \cite{Jia:2009np}.

The LO quark-level subprocesses $e^+e^- \to J/\psi + c+\bar{c}$ and $e^+e^- \to J/\psi + g+g$ yield three-body final states. We label the final-state four-momenta as $p_3=P$, $p_4$, and $p_5$.
For the single-photon exchange process, the differential three-body phase space simply reads
\begin{equation}\label{eq:phasespace}
\int \mathrm{d}\Phi_3 = \frac{s}{2(4\pi)^3}\int_{2\sqrt{r}}^{z_3^+}\mathrm{d}z_3\int_{z_4^-}^{z_4^+}\mathrm{d}z_4.
\end{equation}
We introduce the dimensionless kinematic variables
\begin{equation}
z_i = \frac{2 p_i^0}{\sqrt{s}}, \quad r = \frac{4 m_c^2}{s}.
\end{equation}
with $p_i^0$ representing the energy component of $p_i$.
For the $c\bar{c}$ channel, the upper bound of $z_3$ and the integration limits $z_4^{\pm}$ take the form
\begin{equation}
z_3^+ = 1,\quad 
z_4^{\pm} = \frac{1}{2}\left[
2 - z_3 \pm \frac{\sqrt{\left(z_3^2 - 4r\right)\left(1 - z_3\right)}}{\sqrt{1 - z_3 + r}}
\right].
\end{equation}
While for the $gg$ channel, we have
\begin{equation}
z_3^+ = 1+r,\quad z_4^\pm = \frac{1}{2}\left(2 - z_3 \pm \sqrt{z_3^2 - 4r}\right).
\end{equation}

As stated above, we treat the outgoing angular directions of all final-state particles as quantities independent of $\bm{q}^2$. 
Subject to total three-momentum conservation for the final-state system, the magnitudes of momenta exhibit a proportional relationship:
\begin{equation}\label{eq:pirelations}
\frac{|\bm{p}_3(\bm{q}^2)|}{|\bm{p}_3(0)|} = \frac{|\bm{p}_4(\bm{q}^2)|}{|\bm{p}_4(0)|} = \frac{|\bm{p}_5(\bm{q}^2)|}{|\bm{p}_5(0)|}.
\end{equation}
Energy conservation for the final-state system yields
\begin{equation}\label{eq:zirelations}
\sum_{i=3}^{5} z_i(\bm{q}^2) = \sum_{i=3}^{5} z_i(0) = 2.
\end{equation}

The corresponding on-shell constraints for $J/\psi$ read
\begin{equation}\label{eq:onshellz3}
\begin{aligned}
  &\left[z_3(\bm{q}^2)\right]^2 - \frac{4}{s}\left[\bm{p}_3(\bm{q}^2)\right]^2 = \frac{16E_q^2}{s},\\
  &\left[z_3(0)\right]^2 - \frac{4}{s}\left[\bm{p}_3(0)\right]^2 = 4r.
\end{aligned} 
\end{equation}
The on-shell constraints for $i=4,5$ read
\begin{equation}\label{eq:onshellz4z5}
\begin{aligned}
\left[z_i(\bm{q}^2)\right]^2 - \frac{4}{s}\left|\bm{p}_i(\bm{q}^2)\right|^2
&= \left[z_i(0)\right]^2 - \frac{4}{s}\left|\bm{p}_i(0)\right|^2 \\
&=
\begin{cases}
r, & c\bar{c}\text{ channel},\\
0, & gg\text{ channel}.
\end{cases}
\end{aligned} 
\end{equation}

When relativistic effects controlled by $\bm{q}^2$ are included, the effective invariant mass of the $J/\psi$ becomes $2E_q$, which exceeds the nonrelativistic limit $2m_c$ valid at $\bm{q}^2=0$. Combining all kinematic constraints derived above, we find that the magnitude of each final-state three-momentum $\bm{p}_i(\bm{q}^2)$ scales down proportionally relative to its nonrelativistic value at $\bm{q}^2=0$. Using these identities (Eqs.~\ref{eq:pirelations},\ref{eq:zirelations},\ref{eq:onshellz3},\ref{eq:onshellz4z5}), we can analytically derive the expansion coefficients of $z_i$ to arbitrary orders in $\bm{q}^2$, defined as
\[
z_i^{(n)} \equiv \left.\frac{\partial^n z_i}{\partial (\bm{q}^2)^n}\right|_{\bm{q}^2=0}.
\]
Explicit expressions for the first-order expansion coefficients read
\begin{equation}\label{eq:zi1}
\begin{aligned}
z_3^{(1)} &= \frac{8b (a_4+a_5)}{s}, \quad
z_4^{(1)} = \frac{-8b a_4}{s}, \quad
z_5^{(1)} = \frac{-8b a_5}{s},
\end{aligned}
\end{equation}
where the auxiliary dimensionless quantities $a_i$ and $b$ are defined via
\begin{equation}\label{eq:aib}
\begin{aligned}
a_i &= \frac{2\left[\bm{p}_i(0)\right]^2}{z_i(0)s}, \quad
b = \frac{1}{z_3(0) \left(a_3 + a_4 + a_5\right)}.
\end{aligned}
\end{equation}

Using these identities (Eqs.~\ref{eq:zi1},\ref{eq:aib}), we expand the $\bm{q}$-dependent phase space as
\begin{equation}\label{eq:psexp}
\begin{aligned}
  &\mathrm{d}\Phi_3(\bm{q}^2)=\mathrm{d}\Phi_3^{(0)}\left|\frac{\partial(z_3^{(1)},z_4^{(1)})}{\partial(z_3^{(0)},z_4^{(0)})}\right|\\
  &\quad =\mathrm{d}\Phi_3^{(0)}\left[1+\left(\frac{\partial z_3^{(1)}}{\partial z_3^{(0)}}+\frac{\partial z_4^{(1)}}{\partial z_4^{(0)}}\right)\bm{q}^2 + \mathcal{O}(v^4)\right]\\ 
\end{aligned}
\end{equation}
The $\mathcal{O}(q^2)$ contributions arising from the phase space are suppressed by the $s$ term in the denominators of $z_i^{(1)}$, and are negligible, hence omitted in the subsequent calculations.

Next, we expand the amplitude in Eq.~\ref{eq:amp} in powers of $\bm{q}$ up to order $\mathcal{O}(v^2)$. 
\begin{equation}
\begin{aligned}
&\mathcal{M}(q, E_{q},z_i)\equiv \mathcal{M}\left[e^+e^- \to c\bar{c}(^3S_1^{1}) + X \right] \\
&\quad\quad = \mathcal{M}^{(0)} + \left.\mathcal{M}^{(2)}\right|_{E_{q}\to m_c,z_i\to z_i^{(0)}} + \mathcal{O}(v^4).
\end{aligned}
\end{equation}
where
\begin{equation}\label{eq:M0M2}
\begin{aligned}
&\mathcal{M}^{(0)} = \mathcal{M}(0,m_{c},z_i^{(0)}), \quad \mathcal{M}^{(2)} = \frac{1}{2} q^\alpha q^\beta
\frac{\partial^2 \mathcal{M}}{\partial q^\alpha \partial q^\beta}\\
&\quad + \frac{\bm{q}^2}{2 m_{Q}}
\frac{\partial \left[ \sqrt{\frac{m_{c}}{E_{q}}} \mathcal{M}_{q\to 0} \right]}{\partial E_{q}}
+ \bm{q}^2
\sum_{i=3}^{5}\frac{\partial \mathcal{M}_{q\to 0}}{\partial z_{i}} z_i^{(1)}.
\end{aligned}
\end{equation}
The factor $(m_c/E_q)^{1/2}$ arises from the relativistic normalization of the $c\bar{c}$ state. For the S-wave case, the replacement $q^\alpha q^\beta \to \frac{1}{3} \bm{q}^2 \Pi^{\alpha\beta}$ is applied, where $\Pi^{\alpha\beta} = -g^{\alpha\beta} + \frac{p^\alpha p^\beta}{p^2}$.
Subsequently, the amplitude squared up to order $v^2$ can be written as
\begin{equation}
\begin{aligned}
|\mathcal{M}|^2
&= \mathcal{M}^{(0)} \mathcal{M}^{(0)*}
+ \left(\mathcal{M}^{(0)} \mathcal{M}^{(2)*} + \text{h.c.}\right) \\
&= |\mathcal{M}^{(0)}|^2 + \frac{1}{6} \bm{q}^2 \Pi^{\alpha\beta}
\left[
\frac{\partial^2 \mathcal{M}}{\partial q^\alpha \partial q^\beta} \mathcal{M}^{(0)*}
+ \text{h.c.}
\right] \\
&\quad + \frac{\bm{q}^2}{2 m_{Q}}
\frac{\partial \left(\frac{m_{Q}}{E_{q}} \mathcal{M} \mathcal{M}^{*}\right)}{\partial E_{q}}
+ \bm{q}^2
\sum_{i=3}^{5}\frac{\partial (\mathcal{M} \mathcal{M}^{*})}{\partial z_{i}} z_i^{(1)}.
\label{eq:MM}
\end{aligned}
\end{equation}
The final term in Eq.~(\ref{eq:MM}), originating from the kinematic expansion of $z_{i}$ variables, constitutes our new contribution to three-body quarkonium production.
Neglecting this term leads to inconsistencies in the total relativistic corrections derived from different phase-space integration schemes \footnote{We present detailed numerical results illustrating this point in the subsequent section.}.
Accordingly, this work presents the first fully consistent evaluation of relativistic corrections for three-body quarkonium production within the standard expansion framework.

\section{Numerical results and discussion}\label{sec:results}

\begin{figure}[t]
\centering
\includegraphics[width=0.45\textwidth]{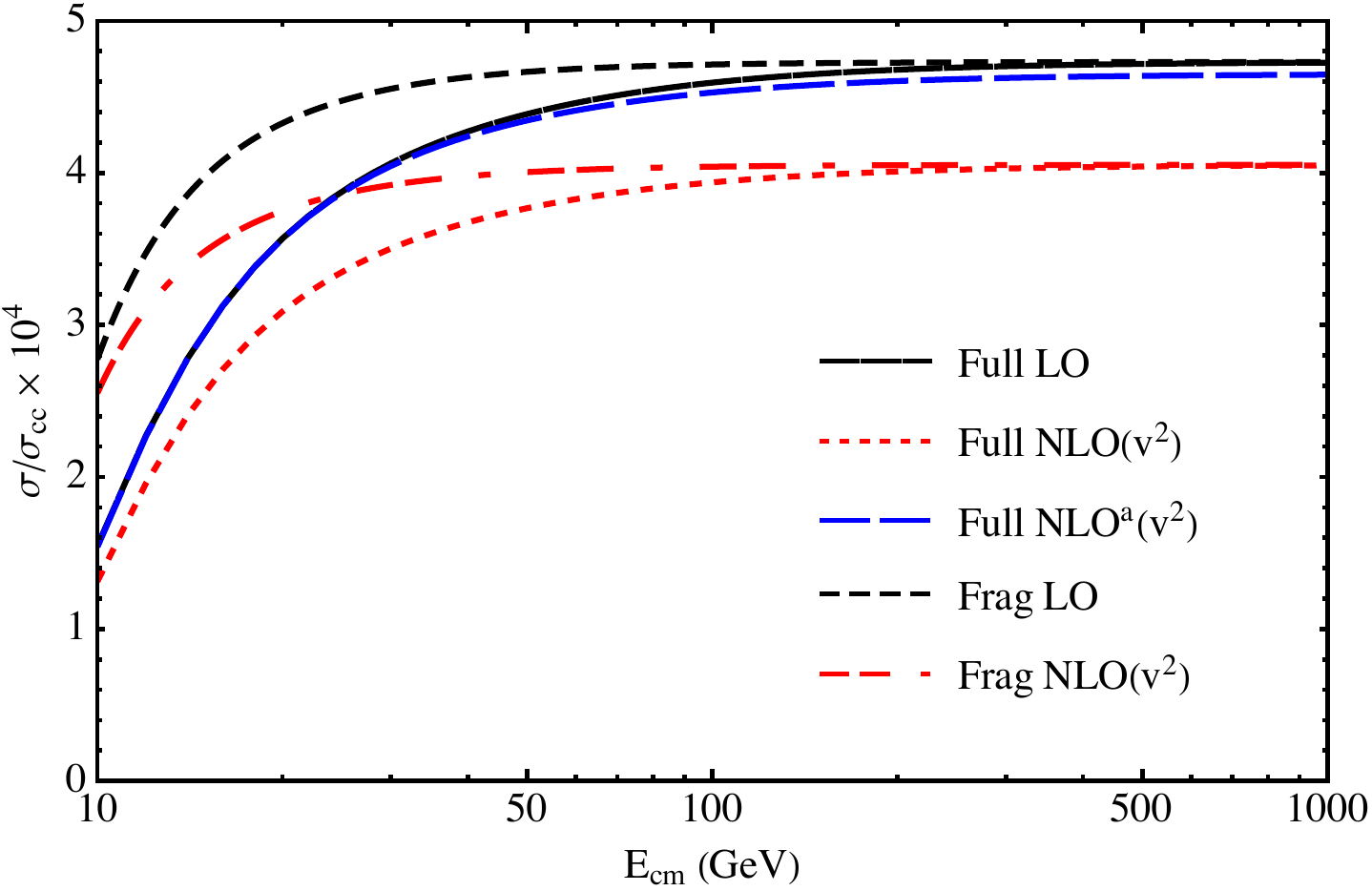}
\caption{Ratios of the cross section $\sigma(e^+e^- \to J/\psi + X_{c\bar{c}})$ to $\sigma(e^+e^- \to c\bar{c})$ as a function of the CM energy $E_{cm}$. "Frag LO" and "Frag NLO" denote the leading-order and next-to-leading-order results in the fragmentation approximation, while "Full LO" and "Full NLO" denote the corresponding results obtained from the full fixed-order calculation.\\
\textsuperscript{a} The results retaining only the contributions form the derivative terms with respect to $q$ and $E_q$.}
\label{fig:jpsicc}
\end{figure}

In the numerical calculations, the LO LDME is evaluated via the radial wave function at the origin:
\begin{equation}\label{eq:wf}
\langle0|\mathcal{O}^{J/\psi}|0\rangle = \frac{9}{2\pi}\left| R_S(0) \right|^2,
\end{equation}
where we take $\left| R_S(0) \right|^2 = 1.01~\mathrm{GeV}^3$ as Refs.~\cite{Zhang:2006ay,Ma:2008gq}.
The relativistic matrix element $\langle v^2 \rangle_{J/\psi}$ is defined as
\begin{equation}\label{eq:v2matrix}
\langle v^2 \rangle_{J/\psi} \equiv \frac{\langle0|\mathcal{P}^{J/\psi}|0\rangle}{m_c^2 \langle0|\mathcal{O}^{J/\psi}|0\rangle},
\end{equation}
with the numerical value $\langle v^2 \rangle_{J/\psi} = 0.23$.
We adopt $m_c = 1.5~\text{GeV}$, $\alpha = 1/137$, and compute the strong coupling $\alpha_s$ via the two-loop running formula with $\Lambda_{\text{QCD}} = 0.388~\mathrm{GeV}$.
The numerical results for two renormalization scales are summarized in Table~\ref{tab:RCC}, where we also present the contributions from $\mathcal{O}(\alpha_s)$ QCD corrections \cite{Zhang:2006ay,Ma:2008gq} for comparison.
For the $J/\psi + X_{\text{non-}c\bar{c}}$ channel, the $\mathcal{O}(v^2)$ correction increases the LO cross section by $14.6\%$, while for the $J/\psi + X_{c\bar{c}}$ channel, the $\mathcal{O}(v^2)$ contribution suppresses the LO yield by $12.2\%$.
Among the $\mathcal{O}(v^2)$ relativistic corrections, the $z_i$ derivative terms newly introduced in this work yield sizable negative contributions for both the $c\bar{c}$ and $\text{non}-c\bar{c}$ channels.
For the $J/\psi + X_{c\bar{c}}$ channel, calculations retaining only derivative terms with respect to $q$ and $E_q$ yield a tiny $\mathcal{O}(v^2)$ correction of roughly $0.003\,\text{pb}$, consistent with the results reported in Ref.~\cite{He:2007te}. For the $J/\psi + X_{\text{non-}c\bar{c}}$ channel, prior studies quote relative corrections of order $20\%-30\%$~\cite{Jia:2009np,He:2009uf} retaining only derivatives with respect to $q$ and $E_q$.
We can find that the correction contributions arising solely from derivatives with respect to $q$ and $E_q$ exhibit scheme dependence with respect to phase-space integration methods.
By contrast, the full relativistic corrections are independent of the phase-space integration scheme.
The newly introduced $z_i$ derivative correction terms are critical for achieving a complete consistent evaluation of relativistic corrections.
This conclusion is further supported by comparisons with the calculations using the fragmentation function.
In Fig.~\ref{fig:jpsicc}, we compare cross-section predictions from our full fixed-order LO and NLO$(v^2)$ calculations against fragmentation results as a function of the CM energy. The fragmentation cross-section formula reads
\begin{equation}
\begin{aligned}
&\sigma_{\text{frag}}(J/\psi + X_{c\bar{c}})
= 2\sigma(e^+ e^- \to c\bar{c})\int_{2\sqrt{r}}^{1} D_{c \rightarrow J/\psi}(z)\, dz .
\end{aligned}
\end{equation}
As illustrated in Fig.~\ref{fig:jpsicc}, our full fixed-order results match fragmentation predictions \cite{Sang:2009zz} remarkably well in the high-energy limit, both at both LO and NLO$(v^2)$\footnote{Notably, fragmentation-function calculations employ a two-body relativistic expansion, with expansions restricted solely to $E_q$ and $q$ in practical computations.}. By contrast, partial calculations that only include derivatives with respect to $q$ and $E_q$ fail to reproduce the fragmentation approximation.
Analogous comparisons between full $\text{NLO}(v^2)$ fixed-order calculations for P-wave quarkonium and fragmentation predictions recover consistent high-energy behavior \cite{Cui:2026bwd}. This further proves our newly derived $z_i$ derivative terms to be mandatory for a fully consistent $\mathcal{O}(v^2)$ treatment of three-body quarkonium production.
Using the same input parameters as in Ref.~\cite{Liu:2003jj} and also taking $\langle v^2 \rangle_{\chi_{cJ}} = 0.23$, we obtain our new $\text{NLO}(v^2)$ $\chi_{c0,1,2} + X_{c\bar{c}}$ cross sections,
\begin{equation}\label{eq:chicjcc}
  \sigma^{\text{NLO}(v^2)}\left(\chi_{c0,1,2} + X_{c\bar{c}}\right)=32.29, 7.33, 4.39\,\text{fb}.
\end{equation}
These cross sections are substantially suppressed relative to the LO predictions, with suppression factors of $0.661$, $0.542$ and $0.697$, respectively.

\begin{table*}[htbp]
\centering
\caption{Cross sections (pb) for different renormalization scales.\footnote{When adopting the phase-space integrand given in Eq.~\ref{eq:phasespace}, the $\mathcal{O}(v^2)$ corrections originating solely from derivatives with respect to $q$ and $E_q$ amount to $-0.692~\text{fb}$ for the $J/\psi + X_{c\bar{c}}$ channels, respectively. Using the phase-space integrand from Refs.~\cite{He:2007te,He:2009uf} instead yields $1.389 ~\text{fb}$, which reproduces the results of Refs.~\cite{He:2007te} under identical input parameters. 
For corrections arising from derivatives of the remaining kinematic parameters, the $\mathcal{O}(v^2)$ contributions $ -25.942~\text{fb}$ with Eq.~\ref{eq:phasespace} integrand, and take the values of $-28.023~\text{fb}$ when employing the integrand of Refs.~\cite{He:2007te}. One can readily confirm that the total correction is invariant under the choice of integration variables.
This feature likewise holds for the $J/\psi + X_{\text{non-}c\bar{c}}$ channel. 
}}
\begin{tabular}{|c|c|ccccc|ccccc|}
\hline
& &
\multicolumn{5}{c|}{$\sigma(J/\psi + X_{\text{non-}c\bar{c}})$} &
\multicolumn{5}{c|}{$\sigma(J/\psi + X_{c\bar{c}})$} \\
\cline{3-12}
$\mu_r$ & $\alpha_s(\mu_r)$ &
LO & $\mathcal{O}(\alpha_s)$ & $\mathcal{O}(v^2)$ & NLO($\alpha_s, v^2$) & $K(\alpha_s, v^2)$ &
LO & $\mathcal{O}(\alpha_s)$ & $\mathcal{O}(v^2)$ & NLO($\alpha_s, v^2$) & $K(\alpha_s, v^2)$ \\
\hline 
$2m_c$ & 0.259 & 0.329 & 0.080 & 0.038 & 0.447 & 1.359 & 0.183 & 0.150 & -0.027 & 0.306 & 1.672 \\
\hline
$\sqrt{s}/2$ & 0.211 & 0.218 & 0.100 & 0.025 & 0.343 & 1.573 & 0.121 & 0.120 & -0.018 & 0.223 & 1.843 \\
\hline
\end{tabular}
\label{tab:RCC}
\end{table*}

\begin{figure*}[t]
\begin{tabular}{cc}
  \includegraphics[width=0.5\textwidth]{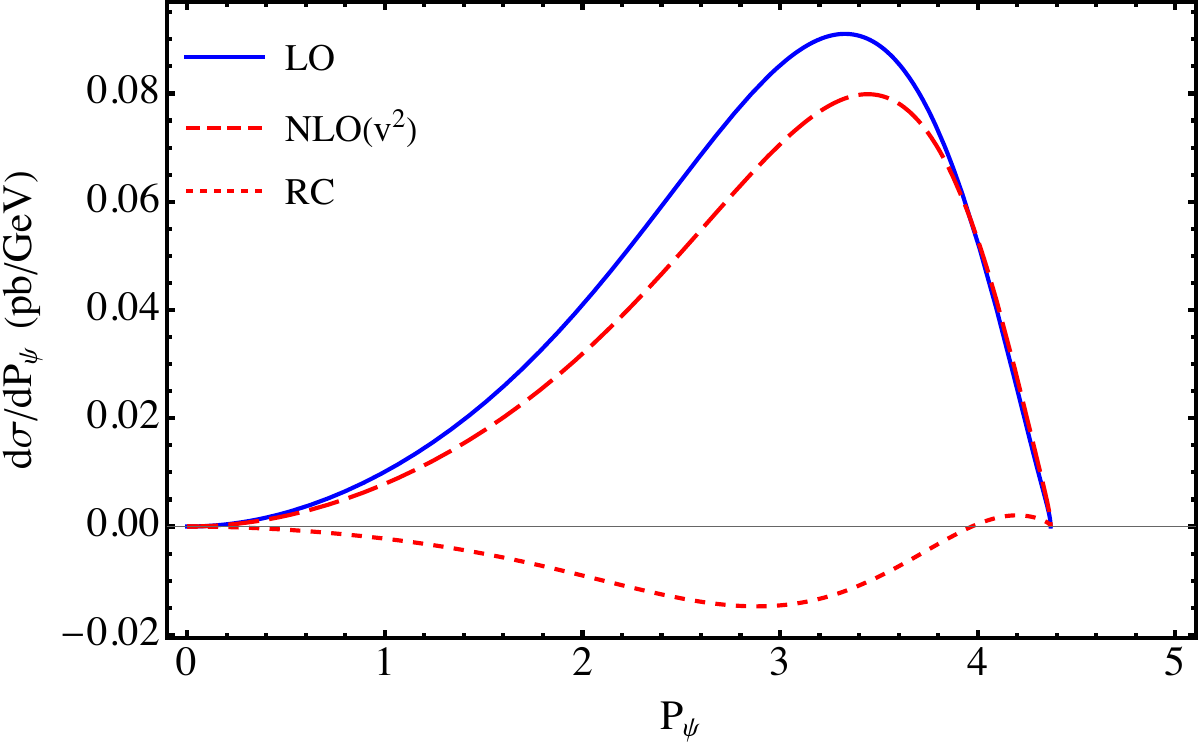}&
  \includegraphics[width=0.5\textwidth]{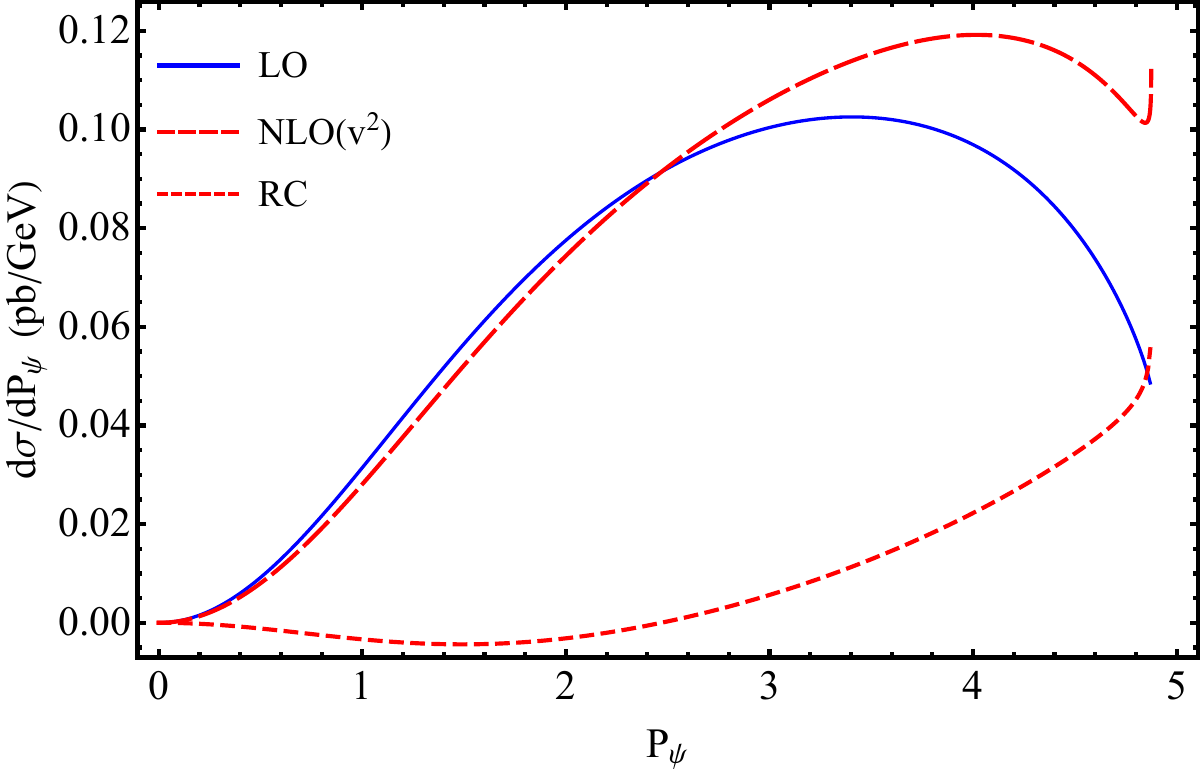}
\end{tabular}
\caption{
the differential cross sections at LO and $\text{NLO}(v^2)$ as functions of the magnitude of the $J/\psi$ three-momentum $P_\psi$. The red short-dashed line labeled RC stands for the result of the $\mathcal{O}(v^2)$ relativistic corrections. The left and right panels correspond to the processes $e^+e^- \to J/\psi+X_{c\bar{c}}$ and $e^+e^- \to J/\psi+X_{\text{non-}c\bar{c}}$, respectively.
\label{fig:ccggpx}}
\end{figure*}

We present the differential cross sections as functions of the $J/\psi$ three-momentum $P_{J/\psi}$ for the two processes, as shown in Fig.~\ref{fig:ccggpx}. 
As visible in the figure, the sign of the $\mathcal{O}(v^2)$ relativistic corrections varies across the full momentum spectrum for both production channels.
For both $J/\psi+X_{c\bar{c}}$ and $J/\psi+X_{\text{non-}c\bar{c}}$, the corrections start negative at low momentum and turn positive at large momentum. The dominant contributions of the $\mathcal{O}(v^2)$ corrections are concentrated in the low-momentum region for the $c\bar{c}$ channel, while for the $\text{non-}c\bar{c}$ channel, sizable positive corrections dominate the high-momentum regime.
The characteristic trend of $\mathcal{O}(v^2)\, c\bar{c}$ curve is also reproduced in fragmentation function calculations \cite{Sang:2009zz}. Analogous to the agreement between full and fragmentation-function cross-section calculations in the high-energy limit (Fig.~\ref{fig:jpsicc}), this feature provides twofold validation for both our full calculations and the corresponding $\mathcal{O}(v^2)$ fragmentation-function results.
The $\mathcal{O}(v^2)$ corrections shift the peak position of the momentum distribution.
To quantify the shift, we introduce the average momentum
$\langle P_{\psi}\rangle$, which is defined as follows:
\begin{equation}
\begin{aligned}
&\langle P_{\psi}\rangle = \frac{\int P_{\psi}\frac{\mathrm{d}\sigma}{\mathrm{d}P_{\psi}}\,\mathrm{d}P_{\psi}}{\int \frac{\mathrm{d}\sigma}{\mathrm{d}P_{\psi}}\,\mathrm{d}P_{\psi}}.
\end{aligned}
\end{equation}
The values listed in Table~\ref{tab:PJPSI} show that peak positions shift to larger momenta for both the $J/\psi+X_{c\bar{c}}$ and $J/\psi+X_{\text{non-}c\bar{c}}$ channels once $\mathcal{O}(v^2)$ corrections are included.
When contrasted against Belle experimental data \cite{Belle:2009bxr}, LO calculations predict peak positions at lower momenta for the $J/\psi+X_{c\bar{c}}$ channel and higher momenta for the $J/\psi+X_{\text{non-}c\bar{c}}$ channel, respectively.
Notably, QCD radiative corrections exert a substantially stronger influence on these peak positions.

\begin{table}[htbp]
\centering
\caption{The average momentum $\langle P_{\psi}\rangle$ at leading order and next-to-leading order in $v^2$.}
\label{tab:PJPSI}
\begin{tabular}{|c|c|c|}
\hline
 & $J/\psi+X_{c\bar{c}}$ & $J/\psi+X_{\text{non-}c\bar{c}}$ \\
\hline
LO & $2.898$ & $2.995$ \\
\hline
NLO($v^2$) & $2.938$ & $3.171$ \\
\hline
\end{tabular}
\end{table}

Following the analysis framework established in Refs.~\cite{Zhang:2006ay,Ma:2008gq}, we include all prompt-production contributions from two-photon channels \cite {Liu:2003zr}, color-octet production \cite{Liu:2003jj,Zhang:2009ym}, and feed-down decays from excited quarkonium states.
Feed-down contributions from $\psi(2S)$ introduce an overall multiplicative factor of $1.355$. The residual contributions to the process $J/\psi+X_{c\bar{c}}+X$ amount to $71~\text{fb}$ as reported in Ref.~\cite{Zhang:2006ay}. We update this value using the below LDMEs. For $J/\psi$ color-octet LDMEs, we utilize global-fit LDMEs for color-octet channels taken from Ref.~\cite{Butenschoen:2011yh}: 
$\langle \mathcal{O}^{J/\psi}(^1S_0^{[8]}) \rangle = 3.04 \times 10^{-2} \, \text{GeV}^3$, $\langle \mathcal{O}^{J/\psi}(^3S_1^{[8]}) \rangle = 1.68 \times 10^{-3} \, \text{GeV}^3$, and $\langle \mathcal{O}^{J/\psi}(^3P_0^{[8]}) \rangle = -9.08 \times 10^{-3} \, \text{GeV}^5$.
These matrix elements exhibit relatively small deviations from measurements performed at $e^+e^-$ colliders \cite{Li:2014fya}.
For the P-wave color-octet LDMEs, we adopt the value $\langle \mathcal{O}^{\chi_{c0}}(^3S_1^{[8]})\rangle = 0.215 \times 10^{-2} \,\text{GeV}^3$ \cite{Ma:2010vd}, and we employ heavy-quark spin symmetry to relate the remaining matrix elements via $\langle\mathcal{O}^{\chi_{cJ}}(^3S_1^{[8]})\rangle = (2J+1)\langle\mathcal{O}^{\chi_{c0}}(^3S_1^{[8]})\rangle$.
Color-octet cross sections are obtained from their color-singlet analogs by rescaling the singlet cross section with the factor $\frac{32 \langle \mathcal{O}_1^H (^{2S+1} L_J)\rangle}{3 \langle\mathcal{O}_8^H (^{2S+1} L_J)\rangle}$.
Finally, the total NLO$(\alpha_s)$ prompt cross sections are expressed as
\begin{equation}
\begin{aligned}
\sigma_{\text{prompt}}^{\text{NLO}(\alpha_s)}(J/\psi+X_{c\bar{c}}+X)
={}&1.355\,\sigma_{\text{direct}}^{\text{NLO}(\alpha_s)}
(J/\psi+X_{c\bar{c}})+49~\text{fb},\\
\sigma_{\text{prompt}}^{\text{NLO}(\alpha_s)}(J/\psi+X_{\text{non}-c\bar{c}})
={}&1.355\sigma_{\text{direct}}^{\text{NLO}(\alpha_s)}
(J/\psi+X_{\text{non}-c\bar{c}})
+321~\text{fb}.
\label{eq:prompt}
\end{aligned}
\end{equation}

Next, we introduce the relativistic corrections. Following previous high-energy-limit analyses \cite{Xu:2012am,Li:2013nna,Xu:2014zra,Wang:2025sbx,Wang:2026gso}, for fragmentation processes of $^3S_1^{[1,8]}$ states, the relativistic correction factor reads $-\frac{11}{6}\langle v^2\rangle$. For the reactions $\gamma^*\to {}^1S_0^{[8]}+g$ and $\gamma^*\to {}^3P_J^{[8]}+g/g^*$, the respective correction factors are $-\frac{5}{6}\langle v^2\rangle$ and $-\frac{31}{30}\langle v^2\rangle$.
For all non-fragmentation $X_{c\bar{c}}$ channels, we recalculate the relativistic corrections using the scheme developed in the present work.
The total NLO$(\alpha_s,v^2)$ prompt cross sections are, therefore, expressed as
\begin{equation}
\begin{aligned}
\sigma_{\text{prompt}}^{\text{NLO}(\alpha_s,v^2)}(J/\psi+X_{c\bar{c}}+X)
={}&1.355\,\sigma_{\text{direct}}^{\text{NLO}(\alpha_s,v^2)}
(J/\psi+X_{c\bar{c}})+29~\text{fb},\\
\sigma_{\text{prompt}}^{\text{NLO}(\alpha_s,v^2)}(J/\psi+X_{\text{non}-c\bar{c}})
={}&1.355
\sigma_{\text{direct}}^{\text{NLO}(\alpha_s,v^2)}
(J/\psi+X_{\text{non}-c\bar{c}})
+259~\text{fb}.
\label{eq:prompt}
\end{aligned}
\end{equation}

\begin{figure*}[t]
\begin{tabular}{cc}
  \includegraphics[width=0.45\textwidth]{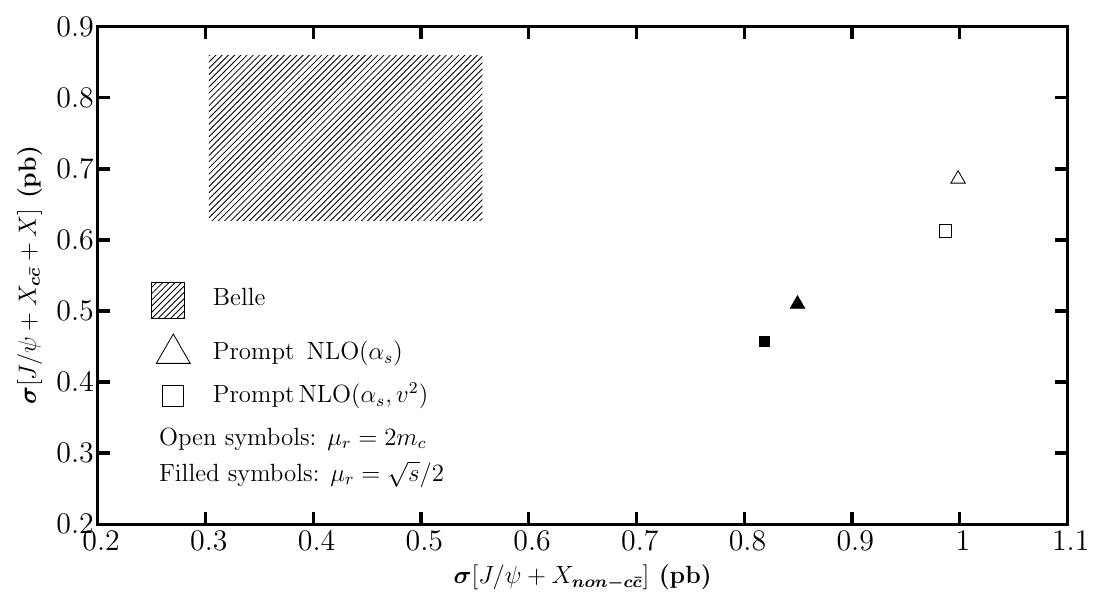}&
  \includegraphics[width=0.45\textwidth]{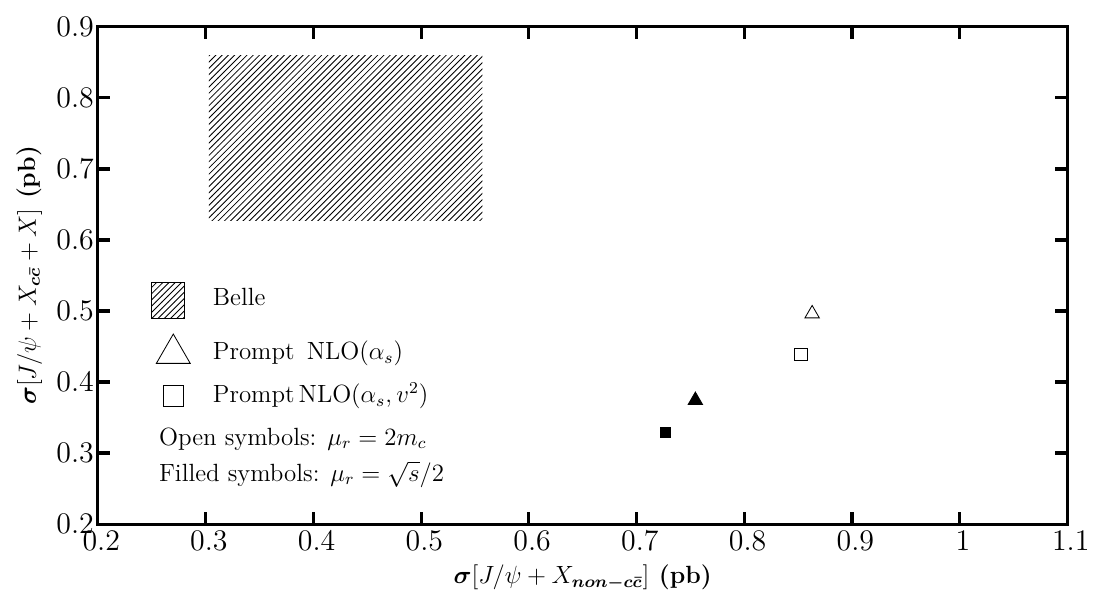}
\end{tabular}
\caption{
Prompt cross sections of the processes $J/\psi+X_{c\bar{c}}+X$ and $J/\psi+X_{\text{non-}c\bar{c}}$ at NLO$(\alpha_s)$ and NLO$(\alpha_s,v^2)$ compared with the Belle measurements. The left and right panels correspond to $m_c=1.4\,\text{GeV}$ and $m_c=1.5\,\text{GeV}$, respectively. 
\label{fig:SGGRCC}}
\end{figure*}

\begin{figure*}[t]
\begin{tabular}{cc}
\includegraphics[width=0.45\textwidth]{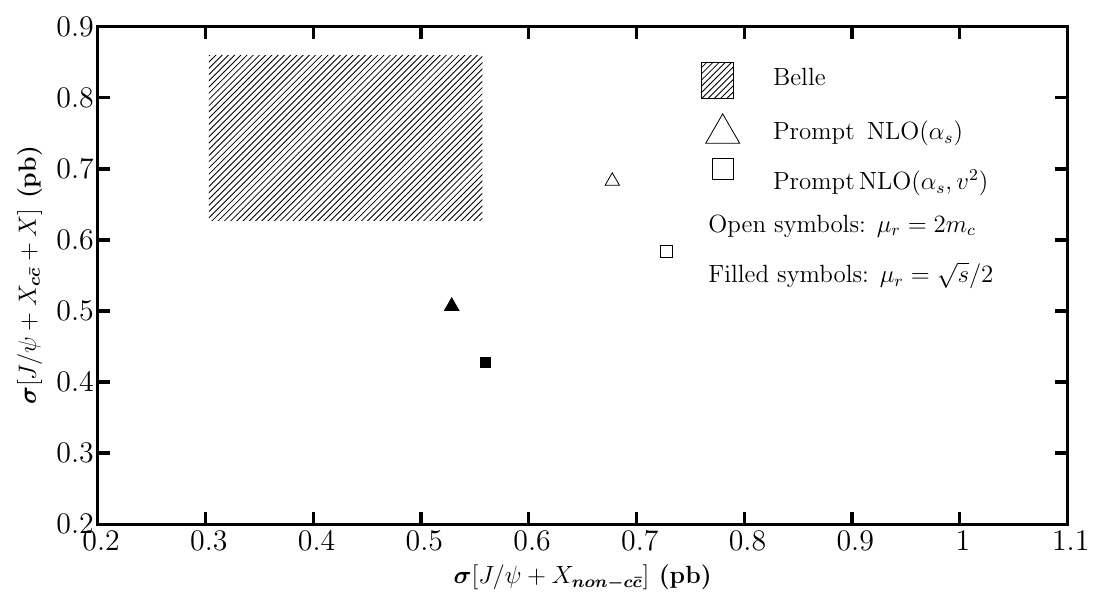}&
\includegraphics[width=0.45\textwidth]{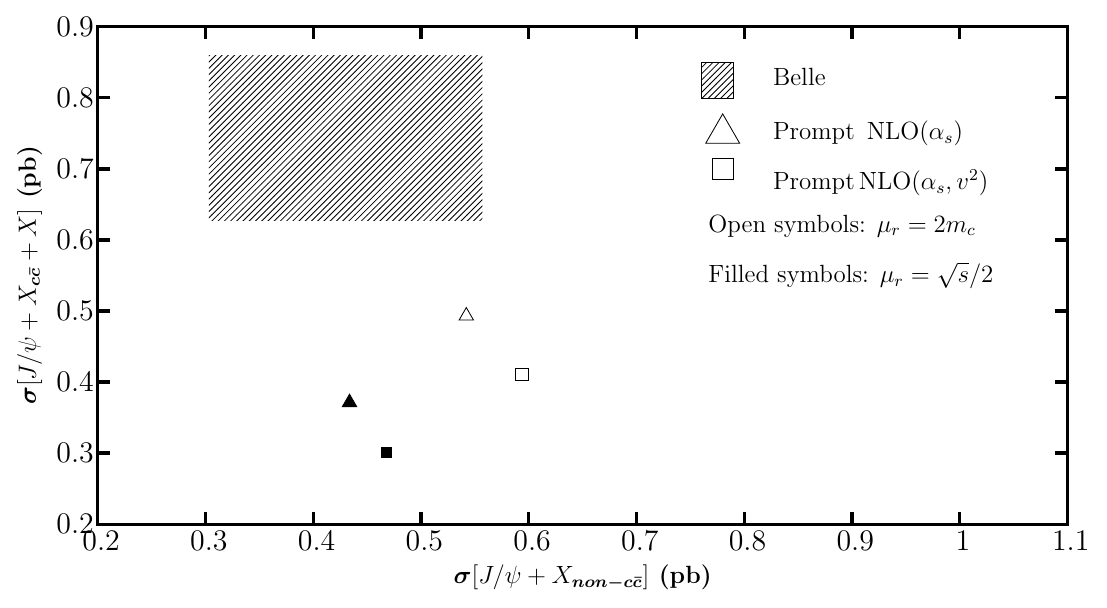}
\end{tabular}
\caption{
Prompt cross sections of the processes $J/\psi+X_{c\bar{c}}+X$ and $J/\psi+X_{\text{non-}c\bar{c}}$ at NLO$(\alpha_s)$ and NLO$(\alpha_s,v^2)$ compared with the Belle measurements. Compared with Fig.~\ref{fig:SGGRCC}, all color-octet contributions are omitted in this figure. The left and right panels correspond to $m_c=1.4\,\text{GeV}$ and $m_c=1.5\,\text{GeV}$, respectively. 
\label{fig:SGGRCCN}}

\end{figure*}

We compare the results with the experimental measurements in Fig.~\ref{fig:SGGRCC}. 
As illustrated in the figures, discrepancies exist between the experimental measurements and both the NLO$(\alpha_s)$ and NLO$(\alpha_s,v^2)$ theoretical predictions. 
One may tune the values of $m_c$ and the renormalization scale can bring the $\mathcal{O}(\alpha_s)$ results for the $X_{c\bar{c}}$ channel into agreement with data, yet the corresponding predictions for the $X_{\text{non-}c\bar{c}}$ channel deviate further from the experimental range. Beyond this, incorporating $\mathcal{O}(v^2)$ relativistic corrections further lowers the predicted cross sections for both channels. 
In theoretical calculations, both color-singlet and color-octet configurations yield comparable cross-section contributions. For the $X_{\text{non-}c\bar{c}}$ channel, color-singlet contributions alone nearly saturate the data, leaving very little room for color-octet channels. 
The figure further demonstrates that the constraints on color-octet long-distance matrix elements extracted from B-factory measurements remain inconsistent with those primarily obtained from hadron collider data. This inconsistency renders the universality of these matrix elements an outstanding open challenge. 
In Fig.~\ref{fig:SGGRCCN}, we present the theoretical predictions when all color-octet contributions are omitted from our calculations. 
Relative to the results shown in Fig.~\ref {fig:SGGRCC}, omitting all color-octet contributions reduces the predicted prompt cross sections for both channels. The discrepancy between the theoretical prediction and experimental data for the $X_{\text{non-}c\bar{c}}$ channel is partially alleviated, while the predicted cross section for the $X_{c\bar{c}}$ channel remains too small. Overall, within the color-singlet framework, new physical mechanisms are still required to account for the residual theory-experiment tension and constrain the allowed room for color-octet contributions.

\section{ SUMMARY}\label{sec:Conclusion}

In this work, we revisit relativistic corrections for the process $e^+e^- \to J/\psi + X$ within the NRQCD factorization framework. 
For quark-level three-body subprocesses, we incorporate new relativistic correction terms derived from expanding all final-state kinematic variables besides $E_q$ (quark–antiquark energy in the meson rest frame) and $q$ (relative four-momentum of the quark–antiquark pair).

For our full calculations of relativistic corrections, the total corrections are invariant under the choice of phase-space integration variables, in contrast to calculations restricted to expansions in $E_q$ and $q$, which show such dependence. The high-energy limit cross sections and line shapes of the energy/momentum distributions for $J/\psi+X_{c\bar{c}}$ agree with fragmentation-function computations \cite{Sang:2009zz} up to $\mathcal{O}(v^2)$.
Complete relativistic expansions are also vital for valid factorization in $\mathcal{O}(v^2)$ computations of P-wave quarkonium production, as seen in inclusive $ h_c$ production in $e^+e^-$ annihilation \cite{Jiang:2026hc}.

We find that the new $\mathcal{O}(v^2)$ cross section for $e^+e^- \to J/\psi+X_{c\bar{c}}$ decreases by 14.55\%, while the cross section for $e^+e^- \to J/\psi+X_{\text{non-}c\bar{c}}$ rises by 12.54\%, compared with previous values $0.4\%$ and $20\%-30\%$, respectively.
After combining published $\mathcal{O}(\alpha_s)$ corrections, two-photon contributions, color-octet channels and feed-down effects from earlier studies, we still observe discrepancies between theoretical predictions and experimental data. A full NLO analysis including both $\mathcal{O}(\alpha_s)$ and $\mathcal{O}(v^2)$ corrections, or higher-order $\mathcal{O}(\alpha_s v^2)$, $\mathcal{O}(\alpha_s^2)$ terms\footnote{Recently, analytical results for $J/\psi+X_{\text{non-}c\bar{c}}$ up to $\mathcal{O}(\alpha_s v^2)$ using the conventional expansion approach are available in Ref.~\cite{Li:2026zsu}, yet they yield negligible contributions to the $\mathcal{O}(v^2)$ and $\mathcal{O}(\alpha_s v^2)$ corrections.}, is expected to mitigate this residual tension.

\section{Acknowledgements:} 
We thank Professor Zhi-Guo He for valuable discussions concerning the present research.
This work was supported by the National Natural Science Foundation of China (No. 11705078, 12575087).

\medskip

%

\end{document}